\documentclass[prd,aps,amsmath,amssymb,preprintnumbers,twocolumn,showpacs,floats]{revtex4}
\input{epsf}
\usepackage{graphics}
\usepackage{amsmath}
\usepackage{amsfonts}
\usepackage{amssymb}
\begin{document}
\title{General form of magnetization damping: Magnetization dynamics of a 
spin system evolving nonadiabatically and out of equilibrium}
\author{F. M. Saradzhev}
\affiliation{Department of Physics, University of Alberta, Edmonton,
Alberta, Canada}
\author{F. C. Khanna}
\affiliation{Department of Physics, University of Alberta, Edmonton,
Alberta, Canada and\\
TRIUMF, 4004 Westbrook Mall, Vancouver, British Columbia, Canada}
\author{Sang Pyo Kim}
\affiliation{Department of Physics, Kunsan National University,
Kunsan 573-701, Korea}
\author{M. de Montigny}
\affiliation{Facult\'e Saint-Jean and Department of Physics,
University of Alberta, Edmonton, Alberta, Canada}

\begin{abstract}

Using an effective Hamiltonian including the Zeeman and 
internal interactions, we describe the quantum theory of magnetization
dynamics when the spin system evolves non-adiabatically and out of
equilibrium. The Lewis-Riesenfeld dynamical invariant method is
employed along with the Liouville-von Neumann equation for the
density matrix. We derive a dynamical equation for magnetization defined
with respect to the density operator with a general form of 
damping that involves the non-equilibrium
contribution in addition to the Landau-Lifshitz-Gilbert equation.
Two special cases of the radiation-spin interaction and the spin-spin 
exchange interaction are considered. For the radiation-spin interaction, 
the damping term is shown to be of the Gilbert type, while in  the 
spin-spin exchange interaction case, the results depend on a coupled chain 
of correlation functions.

\end{abstract}


\maketitle

\section{Introduction}

Magnetization dynamics in nanomagnets and thin films is rich in
content, including such phenomena as giant magnetoresistance
\cite{bai}, spin-current-induced magnetization reversal
\cite{slon},\cite{baza} and adiabatic spin pumping \cite{tser}. The
study of the magnetization dynamics is motivated by theoretical
interest in a deep and fundamental understanding of the physics of
magnetic systems on a short time scale and out of equilibrium.

The results are important in gaining an understanding of the
technological applications of magnetic systems to areas such as
high-density memory and data storage devices. Recent results, in
real time, provide the behavior of spins in particular set-ups of
magnetic fields and measuring the spin flips as a function of time
\cite{myers},\cite{free}.

The Landau-Lifshitz-Gilbert (LLG) equation \cite{dau},\cite{gil}
provides a plausible phenomenological model for many experimental
results. Recently, the LLG equation and the Gilbert damping term
have been derived from an effective Hamiltonian including the
radiation-spin interaction (RSI) \cite{ho}. It has been assumed
there that the spin system maintains quasiadiabatic evolution.

However, a magnetic system whose Hamiltonian $\hat{\cal H}(t)$
evolves nonadiabatically, i.e. in a non-equilibrium state, deviates
far from quasi-equilibrium, and its density operator satisfies the
quantum Liouville - von Neumann (LvN) equation
\begin{equation}
i \hbar \frac{{\partial}\hat{\rho}}{{\partial}t} + [ \hat{\rho},
\hat{\cal H} {]}_{-}=0. \label{llg}
\end{equation}
In the present work, we aim to derive a magnetization equation
including a damping term for such a nonequilibrium magnetic system.
To find its nonadiabatic quantum states, we employ the
Lewis-Riesenfeld (LR) dynamical invariant method \cite{lewis}. This
method originally designed for the nonequilibrium evolution of
time-dependent quantum systems has been successfully applied to a
variety of problems, including the nonadiabatic generalization of
the Berry phase for the spin dynamics, nonequilibrium fermion
systems etc \cite{miz},\cite{san},\cite{sengupta}. In \cite{kim},
the time-dependent invariants have been used for constructing the
density operator for nonequilibrium systems. Following \cite{kim},
we construct the density operator and then use it to determine the
time evolution of magnetization.

The damping term represents magnetization relaxation processes due
to the dissipation of magnetic energy. Various kinds of relaxation
processes are usually melded together into a single damping term.
Relativistic relaxation processes result in the Gilbert damping term
with one damping parameter, while for the case of both exchange and
relativistic relaxation the damping term is a tensor with several
damping parameters \cite{bar}, \cite{saf}.

The relaxation processes are specified by interactions of spins with
each other and with other constituents of the magnetic system. A
derivation of the damping term from first principles should
therefore start with a microscopic description of the interactions.
Even though such microscopic derivation of damping has been
performed for some relaxation processes (for instance, \cite{kam}),
a full version of derivation for the Gilbert damping term has not
yet been given, and in particular for a system in non-equilibrium.

Herein, we consider a general spin system without specifying the
interaction Hamiltonian and related relaxation processes. We start
with a system of spins precessing in the effective magnetic field
$\vec{\bf H}_{eff}$ neglecting for a moment mutual interactions.
Then at a fixed time later interactions in the system are switched
on and influence the original precessional motion.

The interactions are assumed to be time-dependent, and the spin
system evolves nonadiabatically out of equilibrium trying to relax
to a new equilibrium magnetization. We perform a transformation,
which is analogous to the one used in the transition to the
interaction picture, to connect the density and magnetic moment
operators before and after the time, when a new nonequilibrium
dynamics starts, and to find an explicit expression for the
interaction contribution to the magnetization equation.

Our paper is organized as follows. In Sec. II, the magnetization
equation for the system of spins with a general form of interactions
is derived. In Sec. III, the case of the radiation-spin interaction
is considered, which is shown to produce the Gilbert damping term
even for systems that are not in equilibrium. The contribution of
the non-dissipative part of the radiation field to the magnetization
equation and magnetization algebra is discussed. Sec. IV focuses on
a special type of spin-spin interactions. We conclude with
discussions in Sec. V.

\section{Magnetization equation}

Let us consider a quantum spin system defined by
\begin{equation}
\hat{\cal H} = \hat{\cal H}_0 + {\lambda} \hat{\cal H}_{I},
\label{1}
\end{equation}
where $\hat{\cal H}_0$ is the Zeeman Hamiltonian describing the
interaction of spins with an effective magnetic field
\begin{equation}
\hat{\cal H}_0 = - {\gamma} \sum_{i} \hat{S}_{i} \cdot \vec{\bf
H}_{eff}(t), \label{2}
\end{equation}
${\gamma}$ being the gyromagnetic ratio and $\hat{S}_i$ being the
spin operator of the ith atom, while the Hamiltonian $\hat{\cal H}_{I}$ 
describes the internal interactions between the constituents of the spin 
system, including, for instance, the exchange and dipolar interactions 
between the atomic spins, as well as higher-order spin-spin interactions.

In Eq.(\ref{2}), the effective magnetic field
$\vec{\bf H}_{eff}$ is given by the energy variational with
magnetization , $\vec{\bf H}_{eff}=- {\delta}E(M)/{\delta}\vec{\bf
M}$, where $E(M)$ is the free energy of the magnetic system. This
field includes the exchange field, the anisotropy field, and the
demagnetizing field, as well as the external field, $\vec{\bf
H}_{ext}$.

The interaction terms included in $\hat{\cal
H}_{I}$ are in general time-dependent, being switched on
adiabatically or instantly at a fixed time $t_0$. The parameter
${\lambda}$ in (\ref{1}) can be chosen small in order to take into
account the higher order effects perturbatively.

We introduce next the magnetic moment operator
\begin{equation}
\hat{\cal M} \equiv - \frac{{\delta}\hat{\cal H}}{{\delta}\vec{\bf
H}_{ext}}, \label{3}
\end{equation}
which is the response of the spin system to the external field. The
magnetization is defined as an ensemble average of the response
\begin{equation}
\vec{\bf M}=\langle \hat{\cal M} \rangle \equiv \frac{1}{V} {\bf \rm
Tr}\{ \hat{\rho} \hat{\cal M} \}, \label{4}
\end{equation}
where $\hat{\rho}$ is the density operator satisfying the
LvN equation (\ref{llg}) and $V$ is the volume of the system.
The explicit form of the density operator will be shown below.

For systems in equilibrium, the Hamiltonian itself satisfies the LvN
equation and the density operator is expressed in terms of the
Hamiltonian. For nonequilibrium systems, whose Hamiltonians are
explicitly time dependent, the density operator is constructed by
making use of the time-dependent adiabatic invariants.

\subsection{Zeeman precession}

Let us first derive the magnetization equation for the Zeeman
Hamiltonian. If the interactions, $\hat{\cal H}_{I}$, are switched
off, the magnetic moment operator and the magnetization are
\begin{equation}
\hat{\cal M}_0 = - \frac{{\delta}\hat{\cal H}_0}{{\delta}\vec{\bf
H}_{ext}} = {\gamma} \sum_{i} \hat{S}_i \label{6}
\end{equation}
and
\begin{equation}
\vec{\bf M}_0 = {\langle \hat{\cal M}_0 \rangle}_{0} =
\frac{\gamma}{V} \sum_{i} {\bf \rm Tr}\{ \hat{\rho}_{0} \hat{S}_{i}
\}, \label{7}
\end{equation}
respectively, the operators $\hat{\cal M}_{0}^{a}$, $a=1,2,3$,
fulfilling the $SU(2)$ algebra
\begin{equation}
\Big[ \hat{\cal M}_{0}^{a} , \hat{\cal M}_{0}^{b} {\Big]}_{-} =
i{\hbar}{\gamma} {\varepsilon}^{abc} \hat{\cal M}_{0}^{c}, \label{8}
\end{equation}
where the summation over repeated indices is assumed. In
Eq.(\ref{7}), the subscript "0" in the symbol $\langle ...
{\rangle}_{0}$ indicates using of the density operator
$\hat{\rho}_0$, which satisfies the equation
\begin{equation}
i{\hbar} \frac{{\partial}\hat{\rho}_0}{{\partial}t} + [ \hat{\rho}_0
, \hat{\cal H}_0 {]}_{-}=0. \label{9}
\end{equation}

Let $\hat{\cal I}_0(t)$ be a non-trivial Hermitian operator, which
is a dynamical invariant. That is, $\hat{\cal I}_0(t)$ satisfies the
LvN equation
\begin{equation}
\frac{d\hat{\cal I}_0}{dt} \equiv \frac{{\partial}\hat{\cal
I}_0}{{\partial}t} + \frac{1}{i{\hbar}} [ \hat{\cal I}_0 , \hat{\cal
H}_0 {]}_{-}=0.
\end{equation}
As shown in \cite{lewis}, the eigenstates of $\hat{\cal I}_0(t)$ can
be used for evaluating the exact quantum states that are solutions
of the Schr\"odinger equation. The linearity of the LvN equation
allows us to state that any analytic functional of $\hat{\cal
I}_0(t)$ satisfies the LvN equation provided that $\hat{\cal
I}_0(t)$ satisfies the same equation. In particular, we can use
$\hat{\cal I}_0(t)$ to define the density operator $\hat{\rho}_0$ as
\cite{kim}
\begin{equation}
\hat{\rho}_0 (t) = \frac{1}{\cal Z}_0 e^{ - \beta \hat{\cal I}_0
(t)}, \quad {\cal Z}_0 = {\bf Tr} \{ e^{ - \beta \hat{\cal I}_0 (t)}
\}.
\end{equation}
Here $\beta$ is a free parameter and will be identified with the
inverse temperature for the equilibrium system.

The LvN equation for $\hat{\cal I}_0$ is
formally solved by
\begin{equation}
\hat{\cal I}_0(t) = \hat{U}(t,t_0) \hat{\cal I}_0(t_0) \hat{U}(t_0,t),
\label{10}
\end{equation}
where
\begin{equation}
\hat{U}(t_0,t) \equiv T\exp \Big\{ \frac{i}{\hbar} \int_{t_0}^{t}
d{\tau} \hat{\cal H}_0({\tau}) \Big\} \label{11} ,
\end{equation}
and $T$ denotes the time-ordering operator.

As the Zeeman Hamiltonian is linear in spin operators, we can take
$\hat{\cal I}_0(t)$ of the same form \cite{khanna}
\begin{equation}
\hat{\cal I}_0 (t) = \sum_{i} \hat{S}_i \cdot \vec{\bf R}_0 (t),
\label{inv op0}
\end{equation}
where $\vec{\bf R}_0$ is a vector parameter to be determined by a
dynamical equation. Then, Eq.(\ref{9}) becomes
\begin{equation}
\sum_{i} \hat{S}_i \cdot \Biggl( \frac{d \vec{\bf R}_0}{dt} -
{\gamma} \vec{\bf R}_0 \times  \vec{\bf H}_{eff} \Biggr) = 0
\end{equation}
and we obtain an equation for the vector parameter given by
\begin{equation}
\frac{d \vec{\bf R}_0}{dt} = - |\gamma| \vec{\bf R}_0 \times
\vec{\bf H}_{eff}. \label{rr}
\end{equation}
This equation explicitly describes the vector precessing with
respect to the field $\vec{\bf H}_{eff}$. Thus, without loss of
generality, we can identify $\vec{\bf R}_0$ with the magnetization
vector $\vec{\bf M}_0$ (up to a dimensional constant factor) and
Eq.(\ref{rr}) with the equation of motion of magnetization for the
Zeeman Hamiltonian $\hat{\cal H}_0$.

An alternative method to obtain the magnetization equation is just
to differentiate both sides of Eq.(\ref{7}) with respect to time and
use Eq.(\ref{9}). In this way, we arrive at Eq.(\ref{rr}) with
$\vec{\bf R}_0 = \vec{\bf M}_0$.

\subsection{Interactions and damping}

In the case, when the interactions are present, the density operator
for the full Hamiltonian may be written as
\begin{equation}
\hat{\rho} (t) = \frac{1}{\cal Z} e^{ - \beta \hat{\cal I} (t)},
\quad {\cal Z} = {\bf Tr} \{ e^{ - \beta \hat{\cal I} (t)} \},
\label{den op}
\end{equation}
where $\hat{\cal I}$ is the invariant operator for the system. As
the interaction Hamiltonian is in general non-linear in spin
operators, $\hat{\cal I}(t)$ cannot be taken in the form given by
Eq.(\ref{inv op0}). Moreover, the form of $\hat{\cal I}(t)$ can not
be determined without specifying the interactions.

To derive the magnetization equation in this case, we proceed as
follows. We perform, on $\hat{\rho}(t)$, the transformation defined
by the operator (\ref{11}),
\begin{equation}
\hat{\rho} \to \hat{\rho}_{int} \equiv \hat{U}(t_0,t) \hat{\rho}(t)
\hat{U}(t,t_0), \label{13}
\end{equation}
removing the Zeeman interaction. For systems with $\hat{\cal H}_0$
constant in time, the operator $\hat{U}(t_0,t)= \exp\{ (i/{\hbar})
\hat{\cal H}_0 (t-t_0) \}$ leads to the interaction picture, which
proves to be very useful for all forms of interactions since it
distinguishes among the interaction times. For our system with both
$\hat{\cal H}_0$ and $\hat{\cal H}_{I}$ dependent on time, the
operator (\ref{11}) plays the same role, removing the unperturbed
part of the Hamiltonian from the LvN equation.

Substituting Eq.(\ref{13}) into (\ref{llg}), yields
\begin{equation}
i{\hbar} \frac{{\partial}\hat{\rho}_{int}}{{\partial}t} = {\lambda}
\Big[ \hat{\cal H}_{int}, \hat{\rho}_{int} {\Big]}_{-}, \label{14}
\end{equation}
where
\begin{equation}
\hat{\cal H}_{int}(t) \equiv \hat{U}(t_0,t) \hat{\cal H}_{I}(t)
\hat{U}(t,t_0). \label{15}
\end{equation}
The magnetic moment operator and the magnetization become
\begin{equation}
\hat{\cal M}=\hat{\cal M}_0 + \hat{\cal M}_{I}, \label{16}
\end{equation}
where
\begin{equation}
\hat{\cal M}_{I} \equiv - {\lambda} \frac{{\delta} \hat{\cal
H}_{I}}{{\delta} \vec{\bf H}_{ext}}, \label{17}
\end{equation}
and
\begin{equation}
\vec{\bf M} = \frac{1}{V} {\bf \rm Tr} \{ \hat{\rho}_{int} (
\hat{\cal M}_{0,int} + \hat{\cal M}_{I,int} ) \}, \label{18}
\end{equation}
$\hat{\cal M}_{0,int}$ and $\hat{\cal M}_{I,int}$ being related with
$\hat{\cal M}_0$ and $\hat{\cal M}_{I}$, respectively, in the same
way as $\hat{\cal H}_{int}$ is related with $\hat{\cal H}_{I}$.

The operators $\hat{\cal M}_{0,{int}}$, $\hat{\cal M}_{I,{int}}$ are
generally used to calculate the magnetic susceptibility
\cite{white}. Let us show now how these operators determine the time
evolution of magnetization. The evolution in time of $\hat{\cal
M}_{0,{int}}$ is given by the equation
\[
\frac{{\partial}\hat{\cal M}_{0,{int}}}{{\partial}t} =
\frac{i}{\hbar} \hat{U}(t_0,t) \Big[ \hat{\cal H}_0, \hat{\cal M}_0
{\Big]}_{-} \hat{U}(t,t_0)
\]
\begin{equation}
= {\gamma} \hat{\cal M}_{0,{int}} \times \vec{\bf H}_{eff},
\label{19}
\end{equation}
which is analogous to Eq.(\ref{rr}). It describes the magnetization
precessional motion with respect to $\vec{\bf H}_{eff}$.

The equation for $\hat{\cal M}_{I,{int}}$,
\[
\frac{{\partial}\hat{\cal M}_{I,int}}{{\partial}t} =
\frac{i}{\hbar} \hat{U}(t_0,t)  \Big[ \hat{\cal H}_0, 
\hat{\cal 
M}_I
{\Big]}_{-} \hat{U}(t,t_0)
\]
\begin{equation}
+ \hat{U}(t_0,t) \frac{{\partial}\hat{\cal 
M}_{I}}{{\partial}t} \hat{U}(t,t_0)
\end{equation}
describes more complex magnetization dynamics governed by the 
interaction Hamiltonian $\hat{\cal H}_{I}$. In addition, $\hat{\cal 
M}_{I}$ depends on time explicitly. However, this dynamics includes 
the precessional motion as well, sine the interactions induced 
magnetization is a part of the precessing total magnetization 
$\vec{\bf M}$. Introducing
\begin{equation}
\vec{\bf D}_{I} \equiv \frac{i}{\hbar} \Big[ \hat{\cal H}_0 ,
\hat{\cal M}_{I} {\Big]}_{-} - {\gamma} \hat{\cal M}_{I} \times
\vec{\bf H}_{eff} \label{20}
\end{equation}
to represent deviations from the purely precessional motion, we
bring the equation for $\hat{\cal M}_{I,{int}}$ into the following
form
\[
\frac{{\partial}\hat{\cal M}_{I,{int}}}{{\partial}t} = {\gamma}
\hat{\cal M}_{I,{int}} \times \vec{\bf H}_{eff}
\]
\begin{equation}
+ \hat{U}(t_0,t) \Big( \frac{{\partial}\hat{\cal
M}_{I}}{{\partial}t} + \vec{\bf D}_{I} \Big) \hat{U}(t,t_0).
\label{21}
\end{equation}
Taking the time-derivative of $\vec{\bf M}$ given by Eq.(\ref{18}) and 
using Eq.(\ref{14}), we obtain
\[
\frac{d\vec{\bf M}}{dt} = \frac{1}{V} {\bf \rm Tr} \Big\{ 
\hat{\rho}_{int} \Big(
\frac{{\partial}\hat{\cal M}_{0,int}}{{\partial}t} + 
\frac{{\partial}\hat{\cal 
M}_{I,int}}{{\partial}t} \Big) 
\]
\begin{equation}
+ \frac{\lambda}{i{\hbar}} \Big[ \hat{\cal M}_{0,int} + \hat{\cal 
M}_{I,int} , \hat{\cal H}_{int} {\Big]}_{-} \Big\}.
\label{inter}
\end{equation}
Substituting next Eqs.(\ref{19}) and (\ref{21}) into Eq.(\ref{inter}), 
finally yields
\begin{equation}
\frac{d\vec{\bf M}}{dt} = - |{\gamma}| \vec{\bf M} \times \vec{\bf
H}_{eff} + \vec{\bf D}, \label{22}
\end{equation}
where
\begin{equation}
\vec{\bf D} \equiv {\lambda} \langle \frac{1}{i{\hbar}} \Big[
\hat{\cal M}, \hat{\cal H}_{I} {\Big]}_{-} \rangle + \langle
\frac{{\partial}\hat{\cal M}_{I}}{{\partial}t} + \vec{\bf D}_{I}
\rangle. \label{23}
\end{equation}
Therefore, Eq.(\ref{22}) is the magnetization equation for the
system specified by (\ref{1}). This equation is general since it is
derived without specifying $\hat{\cal H}_{I}$. The $\vec{\bf
D}$-term contains all effects that the interactions, $\hat{\cal
H}_{I}$, can have on the magnetization precession, so that
Eq.(\ref{22}) is complete.

The contribution of $\hat{\cal H}_{I}$ to the $\vec{\bf D}$-term in
the magnetization equation can be divided into two parts. One is
proportional to $\langle [ \hat{\cal M} , \hat{\cal H}_{I} {]}_{-}
\rangle$ and is related to the change in the density matrix when the
interactions of $\hat{\cal H}_{I}$ are switched on. The second part
$\langle \frac{{\partial} \hat{\cal M}_{I}}{{\partial}t} + \vec{\bf
D}_{I} \rangle$ originates from the change in the magnetization
itself. Which part of $\vec{\bf D}$ is dominating depends on the
nature of the interactions.

\section{Radiation-spin interaction}

One of the important issues in the study of the magnetization
dynamics is the relaxation phenomena. The magnetization relaxation
mechanism can be introduced by various interactions such as
spin-orbit coupling and two-magnon scattering. In this section, we
calculate the $\vec{\bf D}$-term for the magnetization relaxation
process, which is induced by the RSI.

\subsection{Dissipative radiation field}

The RSI approach is an effective field method, in which contributions to 
the magnetization relaxation are effectively represented by the 
radiation-spin interaction \cite{ho}.  In this method, the damping 
imposed on the precessing magnetization originates from the magnetization 
precessional motion itself. It is assumed that a radiation field is 
induced by the precession and that this field acts back on the 
magnetization producing a dissipative torque.

The Hamiltonian for the RSI is 
\begin{equation}
{\lambda} \hat{\cal H}_{I}= - {\gamma} \sum_{i} \hat{S}_{i} \cdot
\vec{\bf H}_{r}^{d}, \label{24}
\end{equation}
where
\begin{equation}
\vec{\bf H}_{r}^{d} \equiv {\lambda} ( \vec{\bf M} \times \vec{\bf
H}_{eff} - {\alpha} M^2 \vec{\bf H}_{eff}) \label{25}
\end{equation}
is the dissipative part of the effective radiation field, i.e. the part
responsible for the dissipative torque, and $M$ is the magnitude of
magnetization. The parameter ${\alpha}$ will be specified below,
while ${\lambda}$ in Eq.(\ref{1}) is now the radiation parameter.

The RSI contribution to the magnetic moment is
\begin{equation}
\hat{\cal M}_{I,d} = - {\lambda} ({\alpha} M^2 \hat{\cal M}_{0} -
\hat{\cal M}_{0} \times \vec{\bf M} ), \label{26}
\end{equation}
and the total magnetization becomes
\begin{equation}
\vec{\bf M} = (1-{\lambda}{\alpha}M^2) \langle \hat{\cal M}_{0}
\rangle + {\lambda} \langle \hat{\cal M}_{0} \rangle \times \vec{\bf
M}. \label{27}
\end{equation}

It turns out that the vectors $\vec{\bf M}$ and $\langle \hat{\cal
M}_0 \rangle$ are parallel. Indeed, taking the vector product of
both sides of Eq.(\ref{27}) with $\langle \hat{\cal M}_{0} \rangle$,
yields the equation
\[
\langle \hat{\cal M}_{0} \rangle \times \vec{\bf M} = {\lambda}
\langle \hat{\cal M}_{0} \rangle \times \langle \hat{\cal M}_{0}
\rangle \times \vec{\bf M}
\]
\begin{equation}
= {\lambda} {\langle \hat{\cal M}_{0} \rangle}^2 \Big[ (1-{\lambda}
{\alpha}M^2) \langle \hat{\cal M}_{0} \rangle - \vec{\bf M} \Big],
\label{28}
\end{equation}
which is valid only if $\vec{\bf M}$ is parallel to $\langle
\hat{\cal M}_{0} \rangle$, so that
\begin{equation}
\langle \hat{\cal M}_{0} \rangle = \frac{\vec{\bf
M}}{1-{\lambda}{\alpha}M^2}. \label{29}
\end{equation}

A straightforward calculation shows that
\[
\langle \vec{D}_{I} \rangle = - {\lambda} \langle \frac{1}{i{\hbar}}
\Big[ \hat{\cal M}, \hat{\cal H}_{I} {\Big]}_{-} \rangle
\]
\begin{equation}
= - \frac{{\gamma}{\lambda}}{1-{\lambda}{\alpha}M^2} \vec{\bf M}
\times \vec{\bf M} \times \vec{\bf H}_{eff}, \label{30}
\end{equation}
so that the contributions of $\vec{\bf D}_{I}$ and $[ \hat{\cal M},
\hat{H}_{I} {]}_{-}$ to Eq.(\ref{23}) cancel each other. Calculating
next the time derivative of $\hat{\cal M}_{I,d}$ and using
Eq.(\ref{29}), we obtain
\begin{equation}
\vec{\bf D} = {\alpha} \vec{\bf M} \times \frac{d\vec{\bf M}}{dt},
\label{31}
\end{equation}
provided the following relation between the parameters $\alpha$ and
$\lambda$ holds
\begin{equation}
{\alpha} = \frac{\lambda}{1-{\lambda}{\alpha}M^2}. \label{32}
\end{equation}
Therefore, Eq.(\ref{22}) takes the form of the LLG equation,
${\alpha}$ becoming a damping parameter.

The relation given by Eq.(\ref{32}) reflects the origin of the
radiation field. Both the radiation and damping parameters depend on
the magnetization. The dimensionless parameters independent of $M$
are ${\lambda}_0 \equiv {\lambda} M$ and ${\alpha}_0 \equiv {\alpha}
M$ with the relation
\begin{equation}
{\lambda}_0 = \frac{{\alpha}_0}{1+{\alpha}_0^2}.
\end{equation}

The equivalent form of the LLG equation, which is more suitable for
calculations, is
\begin{equation}
\frac{d\vec{\bf M}}{dt} = - |{\gamma}| \vec{\bf M} \times \Big[
(1-{\lambda}{\alpha}M^2) \vec{\bf H}_{eff} + {\lambda}\vec{\bf M}
\times \vec{\bf H}_{eff} \Big]. \label{34}
\end{equation}
Let us assume that $\vec{\bf H}_{eff}$ is a uniform static field in
the $z$-direction, i.e. $\vec{\bf H}_{eff}=(0,0,H_z)$. Then the
magnetization equation becomes, in component form,
\begin{eqnarray}
\frac{d}{dt} M_p^2 & = & -2{\lambda}{\omega}_0 M_z M_p^2,\\
\frac{d}{dt} M_z & = & {\lambda}{\omega}_0 M_p, \label{35-36}
\end{eqnarray}
where ${\omega}_0 \equiv |{\gamma}| H_z$ is the frequency of
magnetization precession and $M_p^2 \equiv M_x^2+M_y^2=M^2-M_z^2$.
These equations are solved exactly by
\begin{equation}
M_z=M \cdot \frac{{\tanh}[{\lambda}_0 {\omega}_0 (t-t_0)] +d}{1+d
\cdot {\tanh}[{\lambda}_0 {\omega}_0 (t-t_0)]} \label{37}
\end{equation}
and
\[
M_p= \frac{M\sqrt{1-d^2}}{{\cosh}[{\lambda}_0 {\omega}_0(t-t_0)]}
\]
\begin{equation}
\cdot \frac{1}{1+d \cdot {\tanh}[{\lambda}_0 {\omega}_0 (t-t_0)]},
\label{38}
\end{equation}
where $d$ stands for the initial condition
\[
d \equiv \frac{M_z(t=t_0)}{M}.
\]
In the limit $t \to \infty$, $M_z$ tends to $M$ and $M_p$ tends to
zero, so that during the relaxation process the magnetization vector
tends to be parallel to the effective magnetic field, and the
relaxation characteristic time ${\tau}$ is $1/({\lambda}_0
{\omega}_0)$.

\subsection{General radiation field}

The RSI approach can be directly used for studying the effect of the real 
radiation-spin interaction in the magnetization relaxation process. In 
that case, one has to distinguish the real radiation field contribution to 
the damping parameter from the effective one. It is therefore important to 
consider possible generalizations of the ansatz given by Eq.(\ref{25}) to 
get a more detailed picture of the RSI.

The dissipative radiation field in Eq.(\ref{25})
is a composite of two fields, which are parallel and perpendicular
to $\vec{\bf H}_{eff}$, respectively. The field parallel to
$\vec{\bf H}_{eff}$ changes only the frequency of the magnetization
precessional motion, while the torque $\vec{\bf H}_{eff} \times
\vec{\bf M}$ introduces a damping effect as well.

The choice of the field $\vec{\bf H}_r^d$ made in Eq.(\ref{25}) is
not a general one. Let us modify it by applying additional fields
and see how the parameters in the magnetization equation are
changed. If we apply an additional field in the direction of
$\vec{\bf H}_{eff}$ and modify $\vec{\bf H}_r^d$ as
\begin{equation}
\vec{\bf H}_r^d \to \vec{\bf H}_r^d - \overline{\lambda}_0 \vec{\bf
H}_{eff}, \label{39}
\end{equation}
where $\overline{\lambda}_0$ is an arbitrary dimensionless
parameter, then this results in a change of the magnitude of the
effective magnetic field and therefore in the change of the
frequency of the magnetization precession as
\begin{equation}
{\omega}_0 \to {\omega} \equiv {\omega}_0 (1-\overline{\lambda}_0 ).
\label{40}
\end{equation}
For $\overline{\lambda}_0 >1$, the magnetization vector turns upside
down and the precession continues in an opposite direction.

The magnetization damping is not affected by the additional field.
Rescaling of $\vec{\bf H}_{eff}$ in $\vec{\bf H}_r^d$ leads to a
change of the radiation parameter as well,
\begin{equation}
{\lambda}_0 \to \frac{1}{1-\overline{\lambda}_0} {\lambda}_0,
\label{lam}
\end{equation}
so that the relaxation characteristic time remains the same.

If we apply an additional torque,
\begin{equation}
\vec{\bf H}_r^d \to \vec{\bf H}_r^d + \overline{\lambda}_0 \vec{\bf
M} \times \vec{\bf H}_{eff}, \label{41}
\end{equation}
then the parameters in the ansatz (\ref{25}) change as follows
\begin{eqnarray}
{\lambda}_0 & \to & {\lambda}_0 +
\overline{\lambda}_0,\\
{\alpha}_0 & \to &  \frac{{\lambda}_0}{{\lambda}_0 +
\overline{\lambda}_0} {\alpha}_0, \label{42-43}
\end{eqnarray}
and the relaxation characteristic time becomes
\begin{equation}
{\tau} \to \overline{\tau} \equiv {\tau} {\Big( 1 +
\frac{\overline{\lambda}_0}{{\lambda}_0} \Big)}^{-1}, \label{44}
\end{equation}
decreasing for $\overline{\lambda}_0 >0$ and increasing for
$-{\lambda}_0 < \overline{\lambda}_0 < 0$. If the additional torque
is stronger than the original one and in the opposite direction,
i.e. $\overline{\lambda}_0 < -{\lambda}_0$, then the magnetization
vector turns over again. For $\overline{\lambda}_0 =-{\lambda}_0$,
the additional and original torques cancel each other, and there is
no damping.
\begin{figure}[t]
\begin{center}
\scalebox{.7}{
\includegraphics{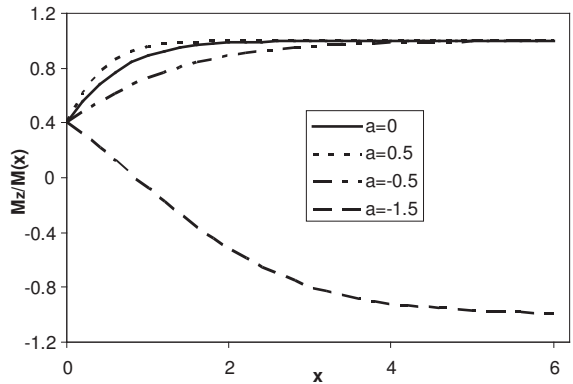}} \caption{The
dependence of the $z$-component of the magnetization vector,
$M_z/M$, on time $x=(t-t_0)/{\tau}$ is shown for different values of
$a$ and for $d=0.4$.} \label{fig1} \end{center}
\end{figure}
\begin{figure}[ht]
\centering \scalebox{.7}{
\includegraphics{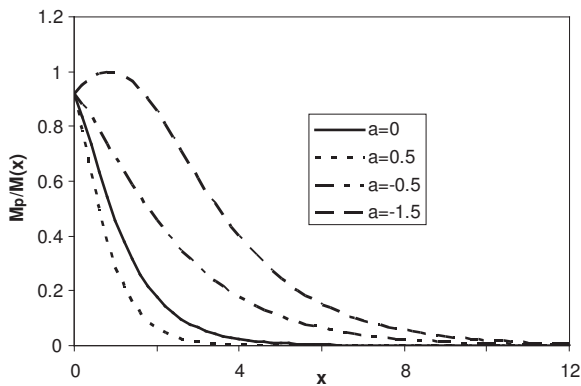}}
\caption{The time dependence of the perpendicular component of the
magnetization vector, $M_p/M$, is plotted for different values of
$a$ and for $d=0.4$.} \label{fig2}
\end{figure}

In Fig.\ref{fig1} and Fig.\ref{fig2} we plot the solutions for $M_z$
and $M_p$ given by Eq.(\ref{37}) and Eq.(\ref{38}), respectively, in
the presence of an additional torque (nonzero $a$'s) and without it
($a=0$), where $a \equiv \overline{\lambda}_{0}/{\lambda}_{0}$. We
observe a change in the relaxation time depending on the sign of
$a$. For $a>0$, the additional torque is in the same direction as
the original one, while for $a<0$ their directions are opposite. A
drastic change occurs for $a<-1$, when $M_z$ tends to $(-M)$
indicating the overturning of $\vec{\bf M}$. The magnetization
vector first becomes perpendicular to $\vec{\bf H}_{eff}$, when
$M_p/M$ reaches its maximum value, and then it turns upside down.

The radiation field $\vec{\bf H}_{r}$ can contain a non-dissipative
part as well, i.e. $\vec{\bf H}_{r} = \vec{\bf H}_{r}^{d} + \vec{\bf
H}_{r}^{n}$. Let us assume that $\vec{\bf H}_r^n$ is parallel to
$\vec{\bf M}$ and we take it of the form
\begin{equation}
\vec{\bf H}_r^n = {\kappa} {\lambda} \frac{\vec{\bf M}}{M} (\vec{\bf
M} \cdot \vec{\bf H}_{eff} ), \label{45}
\end{equation}
where ${\kappa}$ is an arbitrary parameter.

The field given by Eq.(\ref{45}) does not change the magnetization
precessional motion. However, the magnetic moment operator is given
as
\begin{equation}
\hat{\cal M} \to \hat{\cal M} + \hat{\cal M}_{I,n} \equiv \hat{\cal
M} + {\kappa} {\lambda} \frac{\vec{\bf M}}{M} (\hat{\cal M}_0 \cdot
\vec{\bf M}). \label{46}
\end{equation}
The relation between the vectors $\langle \hat{\cal M}_0 \rangle$
and $\vec{\bf M}$ becomes
\begin{equation}
\langle \hat{\cal M}_0 \rangle = \frac{\vec{\bf M}}{1 -
{\lambda}({\alpha}M+{\kappa})M}. \label{47}
\end{equation}

Proceeding in the same way as before, when the non-dissipative part
of the radiation field was omitted, to calculate the total radiation
field contribution to the ${\vec{\bf D}}$-term, we obtain the same
Gilbert-type structure of the damping term with the same damping
parameter, i.e. the part of the radiation field that is parallel to
$\vec{\bf M}$ does not contribute to the magnetization equation.

\subsection{Magnetization algebra}

The magnetization algebra, i.e. the algebra of magnetic moment
operators, gives a further insight into the RSI. Let us study how
the non-dissipative part of the radiation field contributes to this
algebra.

Although we can omit the non-dissipative radiation field in the procedure
obtaining the equation of motion for magnetization, it is important
to take this field into account when we construct the magnetization
algebra. Indeed, the operators $\hat{\cal M}^{a}=\hat{\cal M}^{a}_0
+ \hat{\cal M}^{a}_{I,d}$ without the non-dissipative radiation
field contribution fulfil the algebra
\[
\Big[ \hat{\cal M}^{a} , \hat{\cal M}^{b} {\Big]}_{-} =
i{\hbar}{\gamma} {\varepsilon}^{abc} \{ (1-{\lambda}{\alpha} M^2 )
\cdot \hat{\cal M}^{c}
\]
\begin{equation}
+ \frac{1}{\kappa} {\lambda}M \cdot \hat{\cal M}_{I,n}^{c} \},
\label{48}
\end{equation}
which, however, contains $\hat{\cal M}_{I,n}$ on its right-hand side
and therefore the algebra is not closed.

Let us define, as in Eq.(\ref{46}), the total magnetic moment
operator $\hat{\cal M}_{tot} \equiv \hat{\cal M} + \hat{\cal
M}_{I,n}$. Then the magnetization algebra becomes
\[
\Big[ \hat{\cal M}_{tot}^{a} , \hat{\cal M}_{tot}^{b} {\Big]}_{-} =
i{\hbar}{\gamma} {\varepsilon}^{abc} \{ (1-{\lambda}{\alpha} M^2 +
{\kappa} {\lambda} M ) \cdot \hat{\cal M}^{c}
\]
\begin{equation}
+ ( \frac{1}{\kappa} {\lambda}M - 1 + {\lambda} {\alpha} M^2 ) \cdot
\hat{\cal M}_{I,n}^{c} \}. \label{49}
\end{equation}
Fixing next the parameter ${\kappa}$ by taking it as a solution of
the equation
\begin{equation}
\frac{1}{\kappa} - {\kappa} = \frac{2}{{\alpha}_0}, \label{50}
\end{equation}
that is
\begin{equation}
{\kappa}_{\pm} = - \frac{1}{{\alpha}_0} (1 \mp
\sqrt{1+{\alpha}_0^2}),
\end{equation}
brings the algebra into the closed, standard form for $SU(2)$
algebra
\begin{equation}
\Big[ \hat{\cal M}_{tot}^{a} , \hat{\cal M}_{tot}^{b} {\Big]}_{-} =
i{\hbar}{\gamma}_{tot} {\varepsilon}^{abc} \hat{\cal M}_{tot}^{c},
\label{51}
\end{equation}
where
\begin{equation}
{\gamma}_{tot} \equiv \pm {\gamma} \frac{1}{\sqrt{1+{\alpha}_0^2}}.
\label{par}
\end{equation}
For $0<{\alpha}_0<1$, ${\kappa}_{+}>0$ and ${\kappa}_{-}<0$, the
sign $(+)$ in Eq.(\ref{par}) corresponding to the case
${\kappa}={\kappa}_{+}$ and sign $(-)$ to ${\kappa}={\kappa}_{-}$.

Therefore, the RSI preserves the form of the magnetization algebra
(c.f. Eq.(\ref{8})), by renormalizing the gyromagnetic ratio, only
if the interaction of spins with the non-dissipative part of the
radiation field is included in a proper way. If the direction of
$\vec{\bf H}_r^n$  is the same as that of $\vec{\bf M}$ and
${\kappa}={\kappa}_{+}$, the gyromagnetic ratio remains negative,
while its absolute value decreases. If the direction of $\vec{\bf
H}_r^n$ is opposite to that of $\vec{\bf M}$ and
${\kappa}={\kappa}_{-}$, then the gyromagnetic ratio changes its
sign and becomes positive.

\section{Spin-spin interactions}

The spin-spin interactions among the spins in the system introduce
many body effects, which can be treated perturbatively in the weak
coupling regime. In this case the $\vec{\bf D}$-term can be expanded
in powers of $\lambda$. To demonstrate this, we consider the
spin-spin interactions of a specific type. The interaction between
spins is usually an exchange interaction of the form
\begin{equation}
-2J \sum_{i,j} \hat{S}_i \hat{S}_j = - \frac{2J}{{\gamma}^2}
\hat{\cal M}_0^2, \label{52}
\end{equation}
the coupling constant $J$ being called the exchange integral. We
generalize the ansatz given by Eq.(\ref{52}) by assuming that the
exchange integral depends on the magnetization and introduce the
spin-spin interactions as follows
\begin{equation}
{\lambda}\hat{\cal H}_{I} = \sum_{i,j} J^{ab}(M) \hat{S}_i^a
\hat{S}_j^b, \label{53}
\end{equation}
where $J^{ab}={\lambda}M^aM^b$. Since $\hat{\cal H}_{I}$ does not
depend explicitly on the external field, its contribution to the
magnetic moment operator vanishes, $\hat{\cal M}_{I}=0$.

The non-vanishing commutator $[\hat{\cal M}_0, \hat{\cal H}_{I}
{]}_{-}$ in Eq.(\ref{23}) is the only contribution of the spin-spin
interaction to the magnetization equation, resulting in
\begin{equation}
\vec{\bf D} = \frac{\lambda}{\gamma} \vec{\bf M} \times
\vec{\bf{\Omega}}, \label{54}
\end{equation}
where
\begin{equation}
{\Omega}^a \equiv \langle \Big[ \hat{\cal M}_0^a, \hat{\cal M}_0^b
{\Big]}_{+} \rangle M^b, \label{55}
\end{equation}
and
\begin{equation}
\Big[ \hat{\cal M}_0^a , \hat{\cal M}_0^b {\Big]}_{+} \equiv
\hat{\cal M}_0^a \hat{\cal M}_0^b + \hat{\cal M}_0^b \hat{\cal
M}_0^a. \label{56}
\end{equation}
The correlation function $G^{ab} \equiv \langle \Big[ \hat{\cal
M}_0^a , \hat{\cal M}_0^b {\Big]}_{+} \rangle$ is the sum of spin
correlation functions,
\begin{equation}
G^{ab} = 2{\gamma}^2 \sum_{i} \sum_{j \neq i} \langle \hat{S}_i^a
\hat{S}_j^b \rangle, \label{57}
\end{equation}
excluding the self-interaction of spins. For the standard ansatz
given in Eq.(\ref{52}), $\vec{\bf D}=0$ and the magnetization
equation does not change.

If the spin-spin interactions are turned on at $t=t_0$, so that
$\hat{\rho}(t_0)=\hat{\rho}_0(t_0)$, then, integrating both sides of
Eq.(\ref{14}), we find
\begin{equation}
\hat{\rho}_{\lambda}(t) = \hat{\rho}_0(t_0) +
\frac{\lambda}{i{\hbar}} \int_{t_0}^{t} d{\tau} \Big[ \hat{\cal
H}_{\lambda}(\tau), \hat{\rho}_{\lambda}(\tau) {\Big]}_{-}.
\label{58}
\end{equation}
Substituting Eq.(\ref{58}) into the definition of $G^{ab}$, yields
the equation
\[
G^{ab}(t) = G_0^{ab}(t_0)
\]
\begin{equation}
+ \frac{1}{\gamma} \int_{t_0}^t d{\tau} J^{cd}({\tau}) \Big(
{\varepsilon}^{ace} G^{ebd}({\tau}) + {\varepsilon}^{bce}
G^{aed}({\tau}) \Big), \label{59}
\end{equation}
where
\begin{equation}
G_0^{ab} \equiv \langle \Big[ \hat{\cal M}_0^a , \hat{\cal M}_0^b
{\Big]}_{+} {\rangle}_{0}, \label{60}
\end{equation}
which relates $G^{ab}$ to the third order correlation function, i.e.
the correlation function of the product of three magnetic moment
operators,
\begin{equation}
G^{abc} \equiv \langle \Big[ \Big[ \hat{\cal M}_0^a , \hat{\cal
M}_0^b {\Big]}_{+} , \hat{\cal M}_0^c {\Big]}_{+} \rangle.
\label{61}
\end{equation}
The correlation function $G^{abc}$, in turn, is related to the
fourth order correlation function and etc., and we have therefore an
infinite number of coupled equations for spin correlation functions.
For any practical calculation this infinite hierarchy has to be
truncated. That then defines the approximation scheme which may be
considered on the basis of the physical requirements for the system.
The approximation scheme will depend on the physical properties such
as density and on the strength of the interactions.

If the Hamiltonian $\hat{\cal H}_{I}$ is a small perturbation to the
original $\hat{\cal H}_0$, we can solve Eq.(\ref{58})
perturbatively. In the lowest, zeroth order in $\lambda$, we replace
$\hat{\rho}_{\lambda}(t)$ by $\hat{\rho}_0(t_0)$, so that $G^{ab}
\approx G_0^{ab}(t_0)$. We choose the initial value for $G^{ab}$ as
\begin{equation}
\sum_{i} \sum_{j \neq i} \langle \hat{S}_i^a \hat{S}_j^b {\rangle}_0
= I^{ab} \label{62}
\end{equation}
with $I^{xx}=I^{yy}=0$, $I^{zz}=I$ and $I^{ab}=0$ for $a \neq b$. We
also define again the $z$-direction as the direction of the
effective magnetic field that is chosen uniform and static. Then the
$\vec{\bf D}$-term becomes, in component form,
\begin{eqnarray}
D_x  & = & 2{\lambda} {\gamma} I M_y M_z,\\
D_y  & = & -2{\lambda} {\gamma} I M_x M_z, \\
D_z  & = &  0. \end{eqnarray}
producing two effects in the magnetization equation: the
magnetization is now precessing with respect to $(H_z - 2{\lambda}I
M_z)$, its z-component remaining constant in time, $(d/{dt})M_z=0$,
and the frequency of the precession is
\begin{equation}
\overline{\omega}_{0} \equiv {\omega}_0 \Big( 1- {\lambda}
\frac{2IM_z}{H_z} \Big). \label{64}
\end{equation}
Therefore, in the lowest order of perturbations, when the $\vec{\bf
D}$ is linear in ${\lambda}$, it shifts the direction and the
frequency of the precessional motion without introducing damping
effects. To find a role of the higher powers of ${\lambda}$ in
$\vec{\bf D}$ and to determine how they affect the magnetization
equation, a truncation of the chain of spin correlation equations is
needed. This would require a consistent perturbation approach to the
hierarchy of the coupled equations for the correlation functions.

\section{Discussions}

We have derived a general form of magnetization equation for a
system of spins precessing in an effective magnetic field without
specifying the internal interactions. It can be applied in the study
of magnetization dynamics of any type, including nonequilibrium and
nonlinear effects, provided the interaction of individual spins with
each other and with other degrees of freedom of the system is
specified. For the interactions related to the relaxation processes,
this equation provides a general form of magnetization damping.

This paper uses the dynamical invariant method introduced by Lewis
and Riesenfeld \cite{lewis}. It extends its applicability to
magnetic systems that are in a non-equilibrium state i.e. the
various components in the defining Hamiltonian are time-dependent.
This is in contrast to the earlier attempts to sove a problem for
quasi-adiabatic evolution.

The $\vec{\bf D}$-term in the magnetization equation has been
obtained without using any approximation scheme. It is exact,
accumulates all effects of the internal interactions on the
magnetization precessional motion and can be a starting point for
practical calculations. We have evaluated the $\vec{\bf D}$-term in
two special cases, the RSI and the spin-spin interactions. For the
RSI, which is linear in spin operators, it takes the form of the
Gilbert damping term, the damping and radiation parameters being
interrelated. For the spin-spin interactions, it is determined by
the spin correlation functions, which fulfil an infinite chain of
equations. A further analysis of the $\vec{\bf D}$-term requires an
approximation scheme to truncate the chain in a consistent approach
to higher order calculations.

In our work, we have considered a specific type of the spin-spin
interactions, which do not contribute to the magnetization algebra.
However, if spin-spin interactions depend explicitly on the external
field, the form of the algebra can change. In this case, the total
magnetic moment operator becomes nonlinear in $\hat{\cal M}_0$, and
this results in the magnetization algebra with an infinite chain of
commutation relations. The chain has to be truncated in a way
consistent with the truncation of the chain of equations for the
spin correlation functions in the same approximation scheme.

\acknowledgments

We wish to thank the Natural Sciences and Engineering Research
Council of Canada for financial support. The work of S.P.K is funded
by the Korea Research Foundation under Grant No.
KRF-2003-041-C20053.


\begin{thebibliography}{99}

\bibitem{bai} M.~N.~Baibich, J.~M.~Broto, A.~Fert, F.~ 
Nguyen Van Dau, F.~Petroff, P.~Eitenne, G.~Creuzet, A.~Friederich,
and J.~Chazelas, Phys. Rev. Lett. {\bf 61}, 2472
(1988); P.~M.~Levy, Solid State Phys. {\bf 47}, 367 (1994);
M.~A.~M.~ Gijs and G.~E.~W.~ Bauer, Adv. Phys. {\bf 46}, 285 (1997);
G.~A. ~Prinz, Science {\bf 282}, 1660 (1998).
\bibitem{slon} J.~C.~ Slonczewski, J. Magn. Magn. Mater. {\bf 159},
L1 (1996); {\bf 195}, L261 (1999); L.~ Berger, Phys. Rev. {\bf B
54}, 9353 (1996); J. Appl. Phys. {\bf 89}, 5521 (2001);
\bibitem{baza} Ya.~B.~ Bazaliy, B.~J.~ Jones, and S.~C.~ Zhang, Phys.
Rev. {\bf B 57}, R3213 (1998); {\bf B 69}, 094421 (2004); J. Appl.
Phys. {\bf 89}, 6793 (2001); X.~ Waintal, E.~B.~Myers, P.~W.~ 
Brouwer, and D.~C.~Ralph, Phys. Rev. {\bf B
62}, 12317 (2000); J.~Z.~ Sun, Phys. Rev. {\bf B 62}, 570 (2000);
A.~ Brataas, Yu.~V.~ Nazarov, and G.~E.~W.~ Bauer, Phys. Rev. Lett.
{\bf 84}, 2481 (2000); C.~Heide, P.~E.~ Zilberman, and R.~J.~
Elliott, Phys. Rev. {\bf B 63}, 064424 (2001); S.~Zhang, P.~M.~
Levy, and A.~Fert, Phys. Rev. Lett. {\bf 88}, 236601 (2002); M.~D.~
Stiles and A.~Zangwill, Phys. Rev. {\bf B 66}, 014407 (2002); J.
Appl. Phys. {\bf 91}, 6812 (2002); M.~D.~ Stiles, J.~ Xiao, and A.~
Zangwill, cond-mat/0309289.
\bibitem{tser} Y.~ Tserkovnyak, A.~ Brataas, and G.~E.~W.~ Bauer, Phys.
Rev. Lett. {\bf 88}, 117601 (2002); Phys. Rev. {\bf B 66}, 224403
(2002); {\bf B 67}, 140404 (2003); A.~ Brataas, Y.~Tserkovnyak, 
G.~E.~W.~Bauer, and B.~I.~Halperin, Phys. Rev.
{\bf B 66}, 060404(R) (2002); B.~ Heinrich, Y.~Tserkovnyak, G.~ 
Woltersdorf, A.~Brataas, R.~Urban, and G.~E.~W.~Bauer, Phys. Rev. Lett.
{\bf 90}, 187601 (2003); E.~ \v{S}im\'anek and B.~ Heinrich, Phys.
Rev. {\bf B 67}, 144418 (2003).
\bibitem{myers} M.~ Tsoi, A.~G.~M.~Jansen, J.~Bass, W.-C.~Chiang, 
M.~Seck, V.~Tsoi, and P.~Wyder, Phys. Rev. Lett. {\bf 80}, 4281 
(1998); {\bf 81}, 493(E) (1998); Nature (London) {\bf 406}, 46 (2000);
J.~A.~ Katine, F.~J.~Albert, R.~A.~Buhrman, E.~B.~Myers, and D.~ 
C.~Ralph, Phys. Rev. Lett. {\bf 84}, 3149 (2000); E.~B.~
Myers, D.~C.~Ralph, J.~A.~Katine,  R.~N.~Louie, and R.~A.~ 
Buhrman, Science {\bf 285}, 867 (1999); Phys. Rev. Lett. {\bf
89}, 196801 (2002); F.~J.~ Albert, J.~A.~Katine, R.~A.~ 
Buhrman, D.~C.~Ralph, Appl. Phys. Lett. {\bf
77}, 3809 (2000); Phys. Rev. Lett. {\bf 89}, 226802 (2002); J.~
Grollier, V.~Cros, A.~Hamzic, J.~M.~George, H.~Jaffr\`es, A.~ 
Fert, G.~Faini, J.~Ben~Youssef, and H.~Legall, Appl. Phys. Lett. {\bf 
78}, 3663 (2001); Phys. Rev.
{\bf B 67}, 174402 (2003); J.~E.~ Wegrowe, D. ~Kelly, T.~Truong, 
Ph. ~Guittienne, and J.~-Ph. ~Ansermet, Europhys. Lett.
{\bf 56}, 748 (2001); J. Appl. Phys. {\bf 91}, 6806 (2002); J.~Z.~
Sun, D.~J.~Monsma, T.~S. ~Kuan, M.~J. ~Rooks, D.~W. ~Abraham,
B. ~Oezyilmaz, A.~D. ~Kent, R.~H. ~Koch, 
J. Appl. Phys. {\bf 93}, 6859 (2003); B.~ Oezyilmaz, A.~D.~ 
Kent, D. ~Monsma, J.~Z. ~Sun, M.~J. ~Rooks, and R.~H. ~Koch,
Phys. Rev. Lett. {\bf 91}, 067203 (2003); S.~ Urazhdin, N.~ 
~O. ~Birge, W.~P. ~Pratt, Jr., and J. ~Bass,
Phys. Rev. Lett. {\bf 91}, 146803 (2003); A.~ Barman, V.~V. 
~Kruglyak, R.~J. ~Hicken,  J.~M. ~Rowe, A.~Kundrotaite, J. ~Scott,
and M. ~Rahman, Phys. Rev. {\bf B 69}, 174426 (2004); M.~ Wu,
B.~A. ~Kalinikos, P.~ Krivosik, and C.~E. ~Patton, J. Appl. Phys. {\bf
99}, 013901 (2006); S.~-K.~ Lee, E.~L.~ Hahn, and J.~ Clarke, Phys.
Rev. Lett. {\bf 96}, 257601 (2006).
\bibitem{free} W.~Y.~ Lee, A.~Samad, T.~A. ~Moore, J.~A.~C. ~Bland, 
B.~C.~Choi, Phys. Rev. {\bf B 61}, 6811 (2000);
B.~C.~ Choi, G.~E. ~Ballentine, M. ~Belov, and M.~R. ~Freeman,
Phys. Rev. {\bf B 64}, 144418 (2001); Phys. Rev. Lett.
{\bf 86}, 728 (2001); {\bf 95}, 237211 (2005); X.~ Zhu, Z. ~Liu, V.~ 
Metlushko, P. ~Gr\"utter, M.~R. ~Freeman, Phys. Rev.
{\bf B 71}, 180408 (2005); M.~H.~ Park, Y.~K. ~Hong, B.~C. ~Choi, 
M.~J. ~Donahue, H. ~Han, and S.~H. ~Gee, Phys. Rev. {\bf B 73},
094424 (2006); Q.~F.~ Xiao, J. ~Rudge, B.~C. ~Choi, Y.~K. ~Hong, 
and G. ~Donohoe, Phys. Rev. {\bf B 73}, 104425 (2006);
Z.~ Liu, F. ~Giesen, X. ~Zhu, R.~D. ~Sydora, and 
M.~R. ~Freeman, cond-mat/0606235.
\bibitem{dau} L.~D.~ Landau and E.~M.~ Lifshitz, Phys.
Z. Sowjetunion {\bf 8}, 153 (1935); L.~D.~ Landau, E.~M.~ Lifshitz, and
L.~P.~ Pitaevski, {\it Statistical Physics, Part 2} (Pergamon,
Oxford, 1980), 3rd ed.
\bibitem{gil} T.~L.~ Gilbert, {\it Armor Research Foundation Rep. No. 11}
(Chicago, IL, 1955); Phys. Rev. {\bf 100}, 1243 (1955).
\bibitem{ho} J.~ Ho, F.~C.~ Khanna, and B.~C.~ Choi, Phys. Rev.
Lett. {\bf 92}, 097601 (2004).
\bibitem{lewis} H.~R.~ Lewis~ Jr. and W.~B.~ Riesenfeld, J. Math.
Phys. {\bf 10}, 1458 (1969).
\bibitem{miz} S.~S.~ Mizrahi, Phys. Lett. {\bf A 138}, 465 (1989);
X.~-C.~ Gao, J.~-B.~ Xu, and T.~-Z.~ Qian, Phys. Lett. {\bf A 152},
449 (1991).
\bibitem{san} S.~P.~ Kim, A.~E.~ Santana, and F.~C.~ Khanna, Phys.
Lett. {\bf A 272}, 46 (2000); J.~K.~ Kim and S.~P.~ Kim, J. Phys.
{\bf A 32}, 2711 (1999); S.~P.~ Kim and D.~N.~ Page, Phys. Rev. {\bf
A 64}, 012104 (2001); S.~P.~ Kim, J.~ Korean Phys.~ Soc.~ {\bf 43},
11 (2003); H.~K.~ Kim and S.~P.~ Kim, J. Korean Phys. Soc. {\bf 48}, 119
(2006).
\bibitem{sengupta} S.~P.~ Kim, S.~ Sengupta, and F.~C.~ Khanna,
Phys. Rev. {\bf D 64}, 105026 (2001); S.~P.~ Kim and C.~H.~ Lee,
Phys. Rev. {\bf D 65} 045013 (2002); G.~ Flores-Hidalgo and R.~ O.~
Ramos, Physica {\bf A 326}, 159 (2003).
\bibitem{kim} S.~P.~ Kim and C.~H.~ Lee, Phys. Rev. {\bf D 62},
125020 (2000); S.~P.~ Kim, J. Korean Phys. Soc. {\bf 41}, 643
(2002).
\bibitem{bar} V.~G.~ Baryakhtar, B.~A. ~Ivanov, A.~L. ~Sukstanskii
and E.~Yu. ~Melikhov, Phys. Rev. {\bf B 56}, 619
(1997) and references therein.
\bibitem{saf} V.~L.~ Safonov, J. Magn. Magn. Mater. {\bf 195}, 526
(1999); J. Appl. Phys. {\bf 85}, 4370 (1999); V.~L.~ Safonov and
H.~N.~ Bertram, J. Appl. Phys. {\bf 87}, 5681 (2000); X.~Wang,
H.~N.~ Bertram, and V.~L.~ Safonov, J. Appl. Phys. {\bf 91}, 6920
(2002); {\bf 92}, 2064 (2002).
\bibitem{kam} V.~Kambersky, Canadian J. Phys. {\bf 48}, 2906 (1970);
V.~L.~ Safonov and H.~N.~ Bertram, Phys. Rev. {\bf B 61}, R14893
(2000).
\bibitem{khanna} J.~Ho, F.~C.~ Khanna, and B.~C. Choi, Phys. Rev. {\bf B
70}, 172402 (2004).
\bibitem{white} R.~M.~ White, {\it Quantum Theory
of Magnetism} (Springer-Verlag, Berlin, Heidelberg, New York, 1983),
2nd ed.

\end{thebibliography}
\end{document}